\documentclass[10pt,aps,prb,twocolumn,amsmath,amssymb,superscriptaddress,nobibnotes]{revtex4-1}
\usepackage{amsmath}
\usepackage{graphicx}
\usepackage{color}
\usepackage{multirow}
\usepackage[caption = false]{subfig}
\usepackage{array}
\usepackage[version=4]{mhchem}
\usepackage{mathtools}
\usepackage{siunitx}
\usepackage{centernot}
\usepackage{tikz}
\tikzset{font={\fontsize{11pt}{12}\selectfont}}
\usetikzlibrary{shapes,arrows}
\usepackage{lipsum}

\makeatletter

\renewcommand*{\p@subsection}{}

\renewcommand*{\p@subsubsection}{}
\makeatother

\usepackage{hyperref}
\hypersetup{
	linkcolor=blue,          
	citecolor=blue,        
	urlcolor=blue,           
}

\newcommand{\ket}[1]{|#1\rangle}

\newcommand{\inner}[2]{\langle#1|#2\rangle}

\newcommand{\expecth}[3]{\langle#1|#2|#3\rangle}

\begin{document}
\title{Efficient local energy evaluation for multi-Slater wave functions in orbital space quantum Monte Carlo}

\author{Ankit Mahajan}
\email{ankitmahajan76@gmail.com}
\affiliation{Department of Chemistry, University of Colorado, Boulder, CO 80302, USA}

\author{Sandeep Sharma}
\email{sanshar@gmail.com}
\affiliation{Department of Chemistry, University of Colorado, Boulder, CO 80302, USA}
\begin{abstract}
Recent developments in selected configuration interaction methods have led to increased interest in using multi-Slater trial wave functions in various quantum Monte Carlo (QMC) methods. Here we present an algorithm for calculating the local energy of a multi-Slater wave function in orbital space QMC. For an \textit{ab initio} Hamiltonian, our algorithm has a cost scaling of \(O(n^5 + n_c)\), as opposed to the \(O(n^4n_c)\) scaling of existing orbital space algorithms, where \(n\) is the system size, and \(n_c\) is the number of configurations in the wave function. We present our method using variational Monte Carlo calculations with the Jastrow multi-Slater wave function, although the formalism should be applicable for auxiliary field quantum Monte Carlo. We apply it to polyacetylene and demonstrate the possibility of using a much larger number of configurations than possible using existing methods.

\end{abstract}
\maketitle

\section{Introduction}
Particle-hole excitations from a mean-field reference state have long been used to encode correlation in wave functions.
In single-reference wave function methods, one usually starts from a single Hartree-Fock (HF) configuration, and the electron correlation is built on top of this zeroth-order state by adding particle-hole excitations, often variationally or perturbatively.\cite{szabo2012modern}
When electron interactions are strong, high rank particle-hole excitations from an HF reference are required for a satisfactory description of the electronic structure. It is computationally infeasible to include arbitrarily high-rank excitations for even moderately sized systems due to the rapid increase in their number with system size. Sometimes the strong interactions are confined to a small set of orbitals and electrons termed as the active space.\cite{roos1980complete} For small enough active spaces it is possible to include all possible excitations, or equivalently, all configurations of the Hilbert space, in the variational treatment, and solve the problem exactly in the active space. This is usually only feasible for active spaces smaller than about 20 electrons in 20 orbitals.\cite{vogiatzis2017pushing} Various approximate methods have been devised to stretch this restrictive bound on the active space size.\cite{chan2002highly, BooThoAla-JCP-09, Holmes2016, Motta2017, gidofalvi2008active} Recently, selected configuration interaction (CI) methods, first used about 50 years ago,\cite{bender1969studies,Huron1973} have attracted renewed attention because of their ability to target the most important configurations contributing to the wave function at a relatively small cost.\cite{evangelista2014adaptive,Holmes2016,liu2016ici,scemama2016quantum,tubman2016deterministic} 
Fast implementations of such techniques are now available that allow calculations with millions of configurations on even small work-stations.

%

Due to the remarkable accuracy and easy computability of selected CI wave functions, they have been used in quantum Monte Carlo approaches as trial wave functions. In real space QMC, they can be used as part of Jastrow multi-Slater wave functions, where the Jastrow factors serve to capture the dynamic correlation efficiently, thereby greatly truncating the length of the selected CI expansion required for a given accuracy.\cite{FilUmr-JCP-96,morales2012multideterminant,giner2013using} Multi-Slater wave functions have also been used in the second-quantized setting of auxiliary field QMC (AFQMC),\cite{zhang1995constrained,zhang1997constrained,Motta2017} where the error incurred by the phaseless approximation\cite{ZhaKra-PRL-03} can be controlled by using more accurate trial wave functions.\cite{chang2016auxiliary,landinez2019non} For calculating properties of wave functions in QMC methods, one needs to evaluate the so-called local quantities for a walker in the MC run. For example, for calculating the energy, one averages the local energy given by
\begin{equation}
   E_L[n] = \frac{\expecth{n}{H}{\phi}}{\inner{n}{\phi}} = \sum_I^{n_c} c_I\frac{\expecth{n}{H}{\phi_I}}{\inner{n}{\phi}}
\end{equation}
where \(\ket{\phi} = \sum_I c_I\ket{\phi_I}\) is the multi-Slater wave function, \(\ket{\phi_I}\) being configurations in the CI expansion obtained by particle-hole excitations from a reference configuration, \(n_c\) is the number of configurations, and \(\ket{n}\) is the walker given by electronic positions in real space QMC and by orbital occupations in orbital space QMC. At first glance, this expression seems to suggest that the cost of calculating the local energy for the multi-Slater wave function is \(n_c\) times the cost of calculating it for a single configuration. But this cost can be significantly reduced since CI configurations are not independent, but are obtained by a small set of excitations from the reference. Various algorithms have been proposed to achieve this speed-up in real space QMC,\cite{Clark2011,Filippi2016,scemama2016quantum,Assaraf2017} most efficient being the one recently proposed by Filippi and co-workers.\cite{Filippi2016,Assaraf2017} Their algorithm has a cost scaling of \(O(n^3+n_c)\), where \(n\) is the system size, which has allowed millions of configurations to be included in trial wave functions in real space. No such drastic speed-up has been reported for orbital space algorithms, thus restricting the number of configurations that can be used. In this article, we present an algorithm for calculating the local energy of a multi-Slater wave function that achieves speed-ups analogous to the real space algorithm for orbital space QMC. We will use the variational Monte Carlo (VMC) treatment of Jastrow multi-Slater wave functions to formulate our technique.


In the following, we will start by briefly describing the wave function and defining the required notation. Then we will show how local energy calculations can be sped up by storing some intermediates. Finally, the method is used for polyacetylene ground-state calculations to demonstrate its efficiency.

\section{Theory}
\subsection{Overview}
The Jastrow multi-Slater wave function ansatz is given by
\begin{equation}
  \begin{split}
		\ket{\psi} &= \hat{\mathcal{J}}\ket{\phi},\\
		\hat{\mathcal{J}} &= \exp\left(\sum_{i\geq j}J_{ij}\hat{n}_i\hat{n}_j\right),\\
		\ket{\phi} &= \sum_I c_I \ket{\phi_I},
	\end{split}
\end{equation}
where \(J_{ij}\) are real numbers, \(\hat{n}_i\) is the number operator for spin-orbital \(i\), and \(\ket{\phi_I}\) are electronic configurations forming the multi-configurational state \(\ket{\phi}\). We will suppress spin indices for brevity throughout. Let \(n\) and \(n_c\) be the number of electrons and number of configurations in the CI expansion, respectively. For scaling considerations, we will assume the number of electrons to be proportional to the number of orbitals, serving as a proxy for the system size.

\begin{figure}
	\centering
  \includegraphics[width=0.4\textwidth]{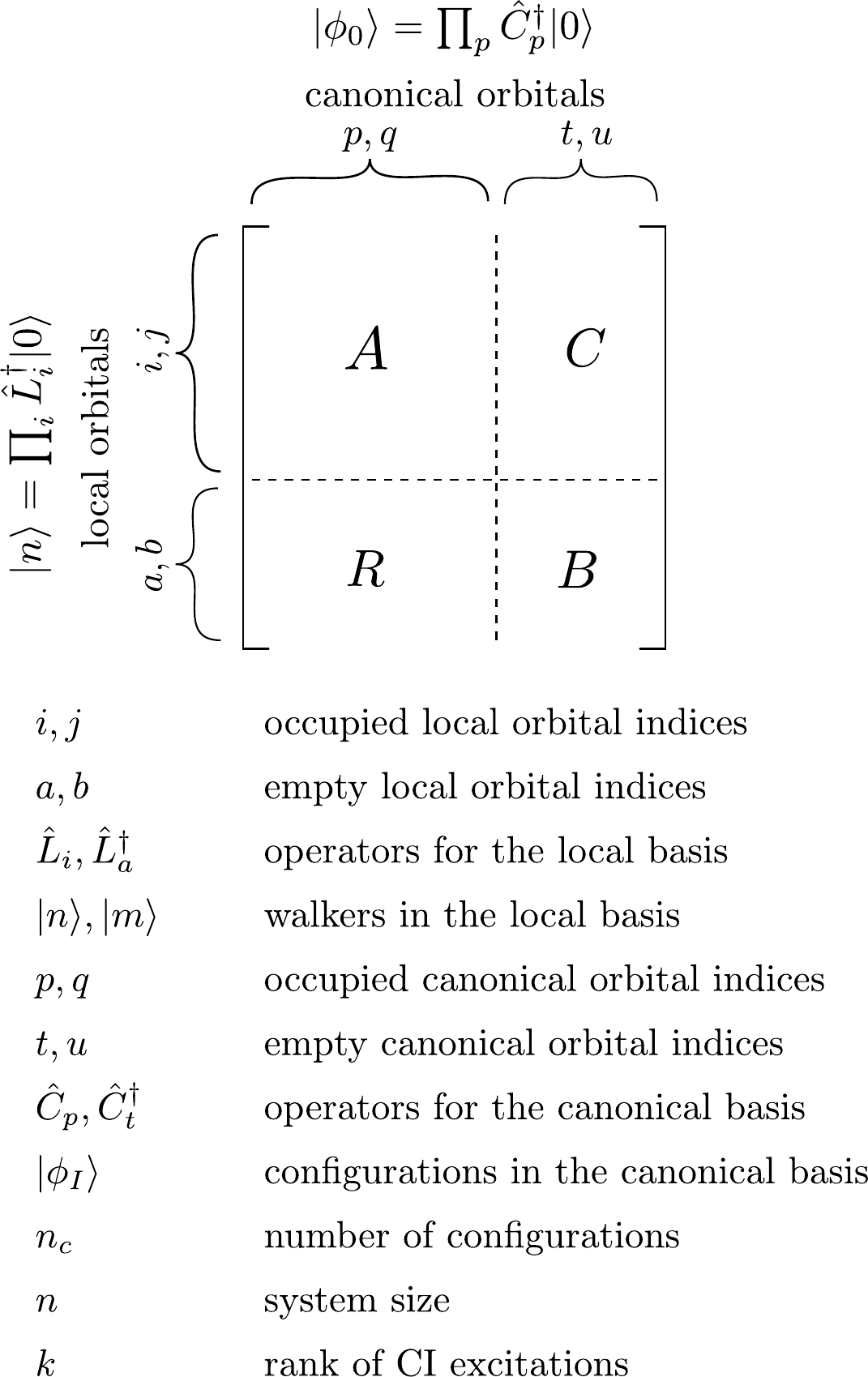}
  \caption{Summary of the notation. The occupied and unoccupied orbitals defining the slices of the coefficient matrix \(M\) are depicted as contiguous here only for clarity, they are interlaced in general.\label{fig:slices}}
\end{figure}

In general, the single-particle orbital sets used for defining the Jastrow and the CI expansion are different.
The second-quantized operators for the spatially local orbitals used in the Jastrow factor\cite{neuscamman2013jastrow} will be denoted as \(\{\hat{L}_{\mu}, \hat{L}_{\mu}^{\dagger}\}\), while those for the canonical orbitals used in the CI expansion as \(\{\hat{C}_{\nu}, \hat{C}_{\nu}^{\dagger}\}\). Indices \(i, j, a, b\) will be reserved for local orbitals, and \(p, q, t, u\) for the canonical ones. We will assume both sets of orbitals to be orthogonal, with the unitary transformation relating them given by
\begin{equation}
   \hat{C}_{\nu} = \sum_i M^{\mu}_{\nu}\hat{L}_{\mu},
\end{equation}
where \(M^i_p\) are real numbers forming the coefficient matrix. We define the following slices of the coefficient matrix:
\begin{equation}
	\begin{split}
		A &= \left[M\right]^{i_1, \dots, i_n}_{p_1, \dots, p_n},\\
		R &= \left[M\right]^{a_1, \dots, a_n}_{p_1, \dots, p_n},\\
		C &= \left[M\right]^{i_1, \dots, i_n}_{t_1, \dots, t_n},\\
		B &= \left[M\right]^{a_1, \dots, a_n}_{t_1, \dots, t_n},\\
	\end{split}
\end{equation}
where superscripts show the row indices and subscripts the column indices used for slicing. Henceforth, all indices will be assumed to be relative to make the notation compact. For example, \(i_{\sigma}\) will label the \(i_{\sigma}\)th occupied local orbital, \(t_{\nu}\) will label the \(t_{\nu}\)th empty canonical orbital, and so on. All configurations in the CI expansion are related to a single configuration \(\ket{\phi_0}\), termed the reference configuration, via a set of canonical orbital excitations:
\begin{equation}
   |\phi_I\rangle = \prod_{\mu=1}^{k_I} E_{\mu}|\phi_0\rangle,
\end{equation}
where \(E_{\mu}\) are single excitation operators. The degree of excitation \(k_I\) for CI configurations is assumed to be of \(O(k)\). In practice, \(k\) is usually less than 10. A summary of the notation is shown in Figure \ref{fig:slices}.

In this article, we are interested in calculating the energy of this wave function, although similar considerations apply to some other physical properties of interest as well. We calculate the energy using variational Monte Carlo sampling:
\begin{equation}
		\frac{\expecth{\psi}{H}{\psi}}{\inner{\psi}{\psi}} = \sum_n \frac{|\inner{n}{\psi}|^2}{\inner{\psi}{\psi}} \frac{\expecth{n}{H}{\psi}}{\inner{n}{\psi}},
\end{equation}
where the walkers \(\ket{n}\) are configurations in the Hilbert space, and the random walk is performed according to the probability distribution \(\inner{\psi}{\psi}\). The quantity being averaged, the local energy, is given by
 \begin{equation}
   \begin{split}
    E_L \left[n\right] &= \frac{\expecth{n}{H}{\psi}}{\inner{n}{\psi}}\\
		&= \sum_m \expecth{n}{H}{m}\frac{\inner{m}{\psi}}{\inner{n}{\psi}},
   \end{split}
\end{equation}
where the configurations \(\ket{m}\) are generated by the action of the Hamiltonian on the walker \(\ket{n}\). The local energy calculation is the rate limiting step in a VMC calculation, and thus efficient algorithms for calculating it are essential for feasible VMC simulations. We use the local orbitals used in the Jastrow to define the walkers. This is a sensible choice because the Jastrow is diagonal in this basis and the Hamiltonian, because of its local interactions, has only quadratically scaling number of two-electron integrals, as opposed to quartic in a general basis. Thus the application of the Jastrow and the Hamiltonian onto a configuration can be performed efficiently in a local basis.\cite{Sabzevari18} This leaves us with the task of calculating the overlap of a walker in a local basis with the multi-Slater wave function defined in a canonical basis. We outline the algorithm for doing this efficiently for local energy calculations in the following sections.

\subsection{Transition matrix elements}
For calculating overlap and local energy, we will need to evaluate the matrix elements of strings of excitation operators between configurations expressed in local and canonical bases. This section summarizes the identities that will be used calculate these transition matrix elements. We will use the generalized Wick's theorem to obtain them.

First note the following pair contractions:
\begin{equation}
	\begin{split}
		\frac{\expecth{n}{\hat{L}_i^{\dagger}\hat{L}_a}{\phi_0}}{\inner{n}{\phi_0}} &= \left[RA^{-1}\right]^a_i,\\
		\frac{\expecth{n}{\hat{C}_t^{\dagger}\hat{C}_p}{\phi_0}}{\inner{n}{\phi_0}} &= \left[A^{-1}C\right]^p_t,\\
		\frac{\expecth{n}{\hat{L}_i^{\dagger}\hat{C}_p}{\phi_0}}{\inner{n}{\phi_0}} &= \left[A^{-1}\right]^p_i,\\
		\frac{\expecth{n}{\hat{L}_a\hat{C}_t^{\dagger}}{\phi_0}}{\inner{n}{\phi_0}} &= \left[B-RA^{-1}C\right]^a_t.
		\label{eq:cont}
	\end{split}
\end{equation}
These can be derived in many different ways.\cite{lowdin1955quantum,balian1969nonunitary,FahyWangLouie88,becca2017quantum,Motta2017} Note that given one of the four identities the rest follow by simply transforming between the \(\hat{L}\) and \(\hat{C}\) operators. Using these pair contractions and the generalized Wick's theorem we get the following transition matrix element
 \begin{equation}
   \begin{split}
		 &\frac{\expecth{n}{\hat{L}_{i_1}^{\dagger}\hat{L}_{a_1}\dots \hat{L}_{i_l}^{\dagger}\hat{L}_{a_l}\hat{C}_{t_k}^{\dagger}\hat{C}_{p_k}\dots \hat{C}_{t_1}^{\dagger}\hat{C}_{p_1}}{\phi_0}}{\inner{n}{\phi_0}} \\
		 &= \det \begin{pmatrix}
		 \left[RA^{-1}\right]^{a_1,\dots,a_l}_{i_1,\dots,i_l} & \left[RA^{-1}C-B\right]^{a_1,\dots,a_l}_{t_1,\dots,t_k}\\[1em]
		\left[A^{-1}\right]^{p_1,\dots,p_k}_{i_1,\dots,i_l}	& \left[A^{-1}C\right]^{p_1,\dots,p_k}_{t_1,\dots,t_k}\\
	 \end{pmatrix},\label{eq:wick}
   \end{split}
\end{equation}
where the RHS is the determinant of a block matrix with blocks given by the indicated slices. The \(RA^{-1}\) elements arise from contractions between \(\hat{L}\) operators, the \(A^{-1}C\) elements from contractions between the \(\hat{C}\) operators, and the \(A^{-1}\) and \(RA^{-1}C-B\) elements from cross contractions between \(\hat{L}\) and \(\hat{C}\) operators. The determinant structure results from the fermionic parity factors. This equation will be used often in the following sections. For convenience, we define the matrix \(\Delta\) as
\begin{equation}
   \Delta = RA^{-1}C-B.
\end{equation}

\subsection{Overlap evaluation}
We will not include the Jastrow factor in the following, since it can be handled the same way as for the Jastrow Slater wave function. The overlap of a walker with the multi-Slater wave function is given by
\begin{equation}
  \inner{n}{\phi} = \sum_I c_I\inner{n}{\phi_I}.
\end{equation}
The first term in this sum, the overlap of the walker with the reference configuration is
\begin{equation}
		\inner{n}{\phi_0} = \expecth{0}{L_{i_n}\dots L_{i_1}C_{p_1}^{\dagger}\dots C_{p_n}^{\dagger}}{0} =\det \left(A\right),
\end{equation}
 where \(\ket{0}\) is the vacuum state. Similarly the overlap of the walker with any other configuration in the CI expansion is also given by a determinant of size \(n \times n\). The cost of calculating these determinants scales as \(O(n^3)\). Thus the cost of calculating the total overlap by evaluating individual determinant overlaps from scratch has a prohibitive cost scaling of \(O(n^3n_c)\). We can improve upon this by storing a few matrices at the start of the calculation.

Suppose \(\ket{\phi_I}\) is obtained from \(\ket{\phi_0}\) by \(k\) excitations as
\begin{equation}
   \ket{\phi_I} = \hat{C}^{\dagger}_{t_k}\hat{C}_{p_k}\dots \hat{C}^{\dagger}_{t_1}\hat{C}_{p_1}\ket{\phi_0}.\label{eq:phiI}
\end{equation}
From equation \ref{eq:wick}, we have
\begin{equation}
	\frac{\inner{n}{\phi_I}}{\inner{n}{\phi_0}} = \det \left(\left[A^{-1}C\right]^{p_1,\dots,p_k}_{t_1,\dots,t_k}\right).
\end{equation}
Thus by calculating the matrix \(A^{-1}C\) and \(\det(A)\) at cost \(O(n^3)\) once, overlap with any configuration in the CI expansion can be calculated at \(O(k^3)\) cost. We note that the matrix \(A^{-1}C\) can be updated using the Sherman-Morrison formula at \(O(n^2)\) cost when the walker makes a move, and does not need to be calculated from scratch every time. Since \(k\) is usually much smaller than \(n\), the total cost scaling of this approach, given by \(O(n^2+n_ck^3)\), proves to be superior in almost all cases.



\subsection{Local energy evaluation}
The local energy of the multi-Slater wave function is given by
\begin{equation}
   \frac{\expecth{n}{H}{\phi}}{\inner{n}{\phi}} = \sum_m\expecth{n}{H}{m}\frac{\inner{m}{\phi}}{\inner{n}{\phi}},\label{eq:eloc}
\end{equation}
where \(\ket{m}\) are generated from the action of the Hamiltonian on the walker \(\ket{n}\). As mentioned before, an \textit{ab initio} Hamiltonian generates \(O(n^4)\) excitations, which is reduced to \(O(n^2)\) when a local basis is used. Because of this large number of excitations, the calculation of local energy constitutes the most demanding part of the VMC calculation. We present two algorithms to evaluate the local energy.

\subsubsection{Algorithm I}
Following the discussion about calculating overlaps, \(\inner{m}{\phi}\) is given by a sum of \(n_c\) determinants. We can again avoid evaluating these determinants individually from scratch by storing a few matrices. Suppose \(\ket{m}\) is obtained from \(\ket{n}\) by \(l\) excitations as
\begin{equation}
   \ket{m} = \hat{L}^{\dagger}_{a_l}\hat{L}_{i_l}\dots \hat{L}^{\dagger}_{a_1}\hat{L}_{i_1}\ket{n}.
\end{equation}
The overlap of \(\ket{m}\) with the excited configuration \(\ket{\phi_I}\) defined in equation \ref{eq:phiI}, is  given precisely by equation \ref{eq:wick} as
\begin{equation}
   \frac{\inner{m}{\phi_I}}{\inner{n}{\phi_0}} = \det \begin{pmatrix}
   \left[RA^{-1}\right]^{a_1,\dots,a_l}_{i_1,\dots,i_l} & \left[\Delta\right]^{a_1,\dots,a_l}_{t_1,\dots,t_k}\\[1em]
	 \left[A^{-1}\right]^{p_1,\dots,p_k}_{i_1,\dots,i_l}	& \left[A^{-1}C\right]^{p_1,\dots,p_k}_{t_1,\dots,t_k}\\
 \end{pmatrix},\label{eq:wick2}
\end{equation}
Thus by calculating \(RA^{-1}\), \(A^{-1}C\) and \(\Delta\) once at cost \(O(n^3)\), each overlap of the form \(\inner{m}{\phi_I}\) can be evaluated at cost \(O(k^3)\), since an \textit{ab initio} Hamiltonian generates only up to double excitations. Note that all of these matrices can be efficiently updated at \(O(n^2)\) cost in a Monte Carlo sampling run.

Having calculated these three matrices, we can directly perform the sum in equation \ref{eq:eloc} to get the local energy:
 \begin{equation}
   \begin{split}
		 E_L[n] &=  \sum_m\expecth{n}{H}{m}\frac{\inner{m}{\phi}}{\inner{n}{\phi}}\\
		 &= \sum_m\sum_Ic_I\expecth{n}{H}{m}\frac{\inner{m}{\phi_I}}{\inner{n}{\phi}},\\
   \end{split}
\end{equation}
where the overlap ratios can be obtained using equation \ref{eq:wick2}. This algorithm has a cost scaling of \(O(n^4n_ck^3)\), if screening of the Hamiltonian elements due to locality of orbitals is not considered, and \(O(n^2n_ck^3)\), if it is. This simple algorithm is easy to implement and allows for incorporating Hamiltonian screening. But the cost scaling is rather steep and one is restricted to fairly short CI expansions. Note that this algorithm is similar to the real space algorithm proposed by Clark \textit{et al.}\cite{Clark2011}. It improves upon their algorithm by obviating the need to calculate extra intermediate matrices.


\subsubsection{Algorithm II}

In the overlap calculation, it was possible to separate the \(n_c\) factor from the system size \(n\) in the cost scaling. Algorithm II achieves this for local energy evaluation by storing some intermediates. First, we partition the local energy based on the rank of the Hamiltonian excitation:
 \begin{equation}
   \begin{split}
       E[n] &= \expecth{n}{H}{n} + \frac{\inner{n}{\phi_0}}{\inner{n}{\phi}}\left(E^{S}[n] + E^{D}[n]\right),\\
			 E^S[n] &= \sum_{ia}H^i_a[n]\frac{\inner{n_i^a}{\phi}}{\inner{n}{\phi_0}},\\
			 E^D[n] &= \sum_{ijab}H^{ij}_{ab}\frac{\inner{n_{ij}^{ab}}{\phi}}{\inner{n}{\phi_0}},\\
   \end{split}
\end{equation}
where
\begin{equation}
  \begin{split}
    \ket{n_i^a} &= \hat{L}_{a}^{\dagger}\hat{L}_{i}\ket{n},\\
		\ket{n_{ij}^{ab}} &= \hat{L}_{b}^{\dagger}\hat{L}_{j}L_{a}^{\dagger}L_{i}\ket{n}
  \end{split}
\end{equation}
are obtained by excitations of the walker through the Hamiltonian. The Hamiltonian matrix elements are defined as \(H^i_a[n] = \expecth{n}{H}{n_i^a}\) and \(H^{ij}_{ab} = \expecth{n}{H}{n_{ij}^{ab}}\). In the following, we will suppress the explicit dependence of certain quantities on the walker whenever it is clear from the context.

Consider the calculation of \(E^S\), arising from single excitations. We can further split it into contributions from individual configurations:
 \begin{equation}
   \begin{split}
       E^S &= \sum_I c_I E^S_I\\
			 E^S_I &= \sum_{ia} H_a^i \frac{\inner{n_i^a}{\phi_I}}{\inner{n}{\phi_0}}.
   \end{split}
\end{equation}
\(E^S_0\) can be calculated directly as in algorithm I:
\begin{equation}
   E^S_0 = \sum_{i,a} H^i_a \left[RA^{-1}\right]^a_i.
\end{equation}
For \(I\neq 0\), suppose \(\ket{\phi_I}\) is obtained from \(\ket{\phi_0}\) by excitations \(\{p_1\rightarrow t_1, \dots, p_k\rightarrow t_k\}\). Using equation \ref{eq:wick2}, \(E^S_I\) can be expressed as
 \begin{equation}
       E^S_I = \sum_{ia} H_a^i \det \begin{pmatrix}
       \left[RA^{-1}\right]^a_i	& \left[\Delta\right]^a_{t_1,\dots,t_k}\\[1em]
			 \left[A^{-1}\right]^{p_1,\dots,p_k}_i & \left[A^{-1}C\right]^{p_1,\dots,p_k}_{t_1,\dots,t_k}\\
		 \end{pmatrix}.
\end{equation}
Note that the determinants appearing in this sum all have the same \(k\times k\) block in the right-bottom corner, they only differ in the first row and column. To see how this can be used to our advantage, let us Laplace expand the determinants along the first column as
 \begin{equation}
   \begin{split}
		 E^S_I &= \det \left(\left[A^{-1}C\right]^{p_1,\dots,p_k}_{t_1,\dots,t_k}\right)\sum_{ia} H^i_a \left[RA^{-1}\right]_i^a\\
		 & + \sum_{\mu=1}^{k}(-1)^{\mu}\det
 \begin{pmatrix}
 \sum_{ia}H^i_a\left[A^{-1}\right]^{p_{\mu}} _i\left[\Delta\right]^{a}_{t_1,\dots,t_k}\\[1em]
 \left[A^{-1}C\right]^{p_1,\dots,p_k\symbol{92} p_{\mu}}_{t_1,\dots,t_k}\\
 \end{pmatrix},
   \end{split}
\end{equation}
where the ordered set of indices \(p_1,\dots,p_k\symbol{92} p_{\mu}\) denotes the set \(p_1,\dots,p_k\) excluding \(p_{\mu}\). In the second line, we have used the linearity of determinants to move the Hamiltonian elements and sum over Hamiltonian excitations inside the determinants. Evidently, the first term is proportional to \(E_0^S\). To efficiently evaluate the second term, let us define the following intermediate:
\begin{equation}
   S^p_t = \sum_{i} \left[A^{-1}\right]^p_i \sum_a \Delta^a_t H^i_a,
\end{equation}
where the sums have been arranged to show the lowest cost scaling order of contractions given by \(O(n^3)\). By building this intermediate during the evaluation of \(E^S_0\), the local energy contribution of configuration \(I\) can be calculated at cost \(O(k^4)\) as
 \begin{equation}
   \begin{split}
		 E^S_I &= \det \left(\left[A^{-1}C\right]^{p_1,\dots,p_k}_{t_1,\dots,t_k}\right)E_0^S\\
		 & + \sum_{\mu=1}^{k}(-1)^{\mu}\det
	 	\begin{pmatrix}
	 	\left[S\right]^{p_{\mu}}_{t_1,\dots,t_k}\\[1em]
	 	\left[A^{-1}C\right]^{p_1,\dots,p_k\symbol{92} p_{\mu}}_{t_1,\dots,t_k}\\
	 	\end{pmatrix}.
   \end{split}
\end{equation}
The total cost scaling of calculating the part of the local energy due to single excitations using algorithm II is \(O(n^3 + n_ck^4)\).  Thus we have managed to seperate the \(n_c\) factor from system size in the cost scaling. We note that this part of the algorithm, dealing with single excitations, is similar to the real space algorithm of Filippi \textit{et al.},\cite{Filippi2016} albeit formulated differently.

Now we turn to the calculation of the double excitation part of the local energy, \(E^D\). Again, we will split it into contributions due to individual configurations:
	\begin{align}
			E^D &= \sum_I c_I E^D_I\\
			E^D_I &= \sum_{ijab} H_{ab}^{ij} \frac{\inner{n_{ij}^{ab}}{\phi_I}}{\inner{n}{\phi_0}}.
	\end{align}
Considering the \(\ket{\phi_I}\) defined above, its contribution is given by
\begin{equation}
			E^D_I = \sum_{ijab} H_{ab}^{ij} \det \begin{pmatrix}
			\left[RA^{-1}\right]^{a,b}_{i,j}	& \left[\Delta\right]^{a,b}_{t_1,\dots,t_k}\\[1em]
			\left[A^{-1}\right]^{p_1,\dots,p_k}_{i,j} & \left[A^{-1}C\right]^{p_1,\dots,p_k}_{t_1,\dots,t_k}\\
		\end{pmatrix}.
\end{equation}
Note that the determinants appearing in this sum all have the same \(k\times k\) block in the right-bottom corner, they only differ in the first \textit{two} rows and columns. To exploit this fact, let us Laplace expand the determinants on the RHS along the first two columns:
\begin{widetext}
\begin{equation}
  \begin{split}
		 E^D_I &= \det \left(\left[A^{-1}C\right]^{p_1,\dots,p_k}_{t_1,\dots,t_k}\right)\sum_{ijab}H^{ij}_{ab}\det
 	   \begin{pmatrix}
     \left[RA^{-1}\right]_{i}^a & \left[RA^{-1}\right]_{j}^a\\[1em]
 		 \left[RA^{-1}\right]_{i}^b & \left[RA^{-1}\right]_{j}^b\\
     \end{pmatrix}\\
 	   &+\sum_{\mu}(-1)^{\mu}\sum_{ijab}H^{ij}_{ab}\det
 	   \begin{pmatrix}
     \left[RA^{-1}\right]_{i}^a & \left[RA^{-1}\right]_{j}^a\\[1em]
 		 \left[A^{-1}\right]_{i}^{p_{\mu}} & \left[A^{-1}\right]_{j}^{p_{\mu}}\\
     \end{pmatrix}
		 \det
		 \begin{pmatrix}
		 \left[\Delta\right]^{b}_{t_1,\dots,t_k}\\[1em]
		 \left[A^{-1}C\right]^{p_1,\dots,p_k\symbol{92} p_{\mu}}_{t_1,\dots,t_k}
		 \end{pmatrix}\\
		 &+\sum_{\mu}(-1)^{\mu+1}\sum_{ijab}H^{ij}_{ab}\det
 	   \begin{pmatrix}
     \left[RA^{-1}\right]_{i}^b & \left[RA^{-1}\right]_{j}^b\\[1em]
 		 \left[A^{-1}\right]_{i}^{p_{\mu}} & \left[A^{-1}\right]_{j}^{p_{\mu}}\\
     \end{pmatrix}
		 \det
		 \begin{pmatrix}
		 \left[\Delta\right]^{a}_{t_1,\dots,t_k}\\[1em]
		 \left[A^{-1}C\right]^{p_1,\dots,p_k\symbol{92} p_{\mu}}_{t_1,\dots,t_k}
		 \end{pmatrix}\\
		 &+\sum_{\mu\nu}(-1)^{\mu+\nu+1}\sum_{ijab}H^{ij}_{ab}\det
		 \begin{pmatrix}
		 \left[A^{-1}\right]_{i}^{p_{\mu}} & \left[A^{-1}\right]_{j}^{p_{\mu}}\\[1em]
	 	 \left[A^{-1}\right]_{i}^{p_{\nu}} & \left[A^{-1}\right]_{j}^{p_{\nu}}\\
		 \end{pmatrix}
		 \det
		 \begin{pmatrix}
		 \left[\Delta\right]^{a,b}_{t_1,\dots,t_k}\\[1em]
		 \left[A^{-1}C\right]^{p_1,\dots,p_k\symbol{92} p_{\mu},p_{\nu}}_{t_1,\dots,t_k}
		 \end{pmatrix}.
  \end{split}
\end{equation}
\end{widetext}
The first term can again be directly related to the local energy contribution of the reference given as
\begin{equation}
   E_0^D = \sum_{ijab}H^{ij}_{ab}\begin{pmatrix}
	      \left[RA^{-1}\right]_{i}^a & \left[RA^{-1}\right]_{j}^a\\[1em]
	  		 \left[RA^{-1}\right]_{i}^b & \left[RA^{-1}\right]_{j}^b\\
	      \end{pmatrix}.
\end{equation}
To efficiently evaluate the second and third terms, we build the intermediates
 \begin{equation}
   \begin{split}
		 \left[D_1\right]^{p}_{t} &= \sum_{ijab}H^{ij}_{ab}\det
 \begin{pmatrix}
 \left[RA^{-1}\right]_{i}^{a} & \left[RA^{-1}\right]_{j}^{a}\\[1em]
 \left[A^{-1}\right]_{i}^{p} & \left[A^{-1}\right]_{j}^{p}\\
 \end{pmatrix}
 \Delta_{t}^{b},\\
 \left[D_2\right]^{p}_{t} &= \sum_{ijab}H^{ij}_{ab}\det
	 \begin{pmatrix}
	 \left[RA^{-1}\right]_{i}^{a} & \left[RA^{-1}\right]_{j}^{a}\\[1em]
	 \left[A^{-1}\right]_{i}^{p} & \left[A^{-1}\right]_{j}^{p}\\
	 \end{pmatrix}
	 \Delta_{t}^{b}.
   \end{split}
\end{equation}
Both of these can built at cost \(O(n^4)\) during the evaluation of \(E_0^D\). For example, one can see how \(D_1\) can be calculated at this cost, since it involves contractions of the type
\begin{equation}
   \sum_j\left[A^{-1}\right]_{j}^{p}\sum_b\Delta_{t}^b\sum_{ia}\left[RA^{-1}\right]_{i}^{a} H^{ij}_{ab},
\end{equation}
obtained by expanding the determinant. Using \(D_1\) and \(D_2\), the second and third terms can be calculated as before due to linearity of determinants at cost \(O(k^4)\). To calculate the final term, we use the intermediate
\begin{equation}
	\begin{split}
	\left[D_3\right]^{pq}_{tu} = \sum_{ijab}H^{ij}_{ab}\Bigg[\Bigg.\det&
	\begin{pmatrix}
	\left[A^{-1}\right]_{i}^{p} & \left[A^{-1}\right]_{j}^{p}\\[1em]
	\left[A^{-1}\right]_{i}^{q} & \left[A^{-1}\right]_{j}^{q}\\
	\end{pmatrix}\\[0.5em]
	&\times\det\left.
	\begin{pmatrix}
	\Delta_{t}^{a} & \Delta_{u}^{a}\\[1em]
	\Delta_{t}^{b} & \Delta_{u}^{b}\\
	\end{pmatrix}\right].
\end{split}\label{eq:d3}
\end{equation}
This intermediate can be built at cost \(O(n^5)\), since it involves sums like
 \begin{equation}
    \sum_i\left[A^{-1}\right]_{i}^{p}\sum_j\left[A^{-1}\right]_{j}^{q}\sum_{a}\Delta_{t}^{a}\sum_{b}\Delta_{u}^{b}H^{ij}_{ab},
 \end{equation}
obtained by expanding the two determinants. Using \(D_3\), the final term can be calculated at cost \(O(k^6)\). Therefore the total cost scaling of the local energy calculation using algorithm II is \(O(n^5 + n_ck^6)\).

Unlike algorithm I, there is no obvious way to use the screening of Hamiltonian elements to reduce the cost scaling of building the intermediates in algorithm II. But we note that the equation for building the intermediate \(D_3\) resembles a Hamiltonian integral transformation. Techniques like density fitting\cite{werner2003fast} and Cholesky decomposition\cite{koch2003reduced} are employed to reduce the cost of such transformations. Tensor hypercontraction\cite{hohenstein2012tensor} can be used to reduce the cost scaling to \(O(n^4)\). Another possible way of improving the cost is by updating the intermediates as the walker moves during a Monte Carlo run, instead of building them from scratch every time. Incorporating such techniques into our algorithm will be a task for future research.

We note that, with some modifications, these algorithms should be applicable for calculating local energies of multi-Slater trial wave functions in AFQMC as well. In AFQMC, three different single-particle basis sets need to be considered: one for the walker, one for the Hamiltonian, and one for the CI expansion.\cite{Motta2017} In this case, the generalized Wick's theorem expression in equation \ref{eq:wick} gets modified only in the first \(l\) rows and columns. The \(k\times k\) block in the right-bottom corner, arising due to contractions between CI excitation operators, remains unchanged. This block allowed us to use the intermediates above, and it should be possible to use them even when the Hamiltonian is expressed in a basis other than the walker basis. Also note that this consideration does not change the scaling of algorithm II, because a Hamiltonian integral transformation is involved in forming the intermediate \(D_3\) regardless. But it would reduce the cost prefactor.

Finally, we point out that the Jastrow factor overlap ratios, that we have ignored in the discussion above, can be absorbed into the intermediates exactly like the Hamiltonian matrix elements.

\subsection{Gradients}
In order to optimize the wave function energy, we use gradient based methods. The \(i\)th component of the energy gradient is sampled according to the equation
\begin{equation}
   \partial_i E = 2 \sum_n\frac{|\inner{n}{\psi}|^2}{\inner{\psi}{\psi}}\frac{\inner{\psi_i}{n}}{\inner{\psi}{n}}\left(E_L[n]-E\right),
\end{equation}
where \(E\) is the energy of the wave function \(\ket{\psi}\), and \(\ket{\psi_i}\) is the derivative of the wave function with respect to the \(i\)th parameter. Since \(E_L[n]\) and \(E\) are both available from the energy sampling, we only need to calculate the wave function derivative overlaps to obtain an estimate of the energy gradient.

The Jastrow multi-Slater wave function has three types of parameters that can be optimized: Jastrow elements, CI coefficients, and orbital coefficients. We note that our parameterization is fairly redundant, and it is possible that removing these redundancies can lead to improvements in optimization. We have chosen not to do so because of the simplicity of the redundant parametrization. For Jastrow parameters, the derivative overlap ratio is given by
\begin{equation}
\frac{\inner{n}{\psi_{J_{ij}}}}{\inner{n}{\psi}} = \frac{n_in_j}{J_{ij}},
\end{equation}
where \(n_i\) and \(n_j\) are occupation numbers of orbitals \(i\) and \(j\) in the walker. All Jastrow derivative overlaps can be calculated at cost \(O(n^2)\).
For the CI coefficients, the derivate overlap ratio is given as
\begin{equation}
\frac{\inner{n}{\psi_{c_{I}}}}{\inner{n}{\psi}} = \frac{\inner{n}{\phi_I}}{\inner{n}{\phi}}.
\end{equation}
All CI coefficient derivative overlaps can be calculated at cost \(O(n_ck^3)\), given the matrices stored during energy sampling. A naive calculation of the orbital coefficient derivatives has a prohibitive cost of \(O(n^2n_ck^3)\), but this can be improved to \(O(n^3+n_ck^3)\) using the method outlined by Assaraf \textit{et al.}\cite{Assaraf2017} We will not make use of orbital optimization in this article.

\subsection{Sampling}
We use continuous time Monte Carlo (CTMC)\cite{Bortz1975,GILLESPIE1976403} to perform the energy and gradient sampling. We refer the reader to our previous work\cite{Sabzevari18} for the details of this sampling technique. To perform efficient CTMC sampling of a wave function \(\ket{\psi}\), one needs to calculate the overlap ratios \(\frac{\inner{m}{\psi}}{\inner{n}{\psi}}\) for all the excitations \(\ket{m}\) generated from the walker \(\ket{n}\) through the Hamiltonian. In algorithm I, all these overlap ratios are calculated directly and can be stored during the local energy calculation, allowing CTMC sampling of the full wave function at essentially no extra cost. But in algorithm II, these overlap ratios are not calculated explicitly, instead they get absorbed in the intermediates. Thus it is not possible to perform efficient CTMC sampling of the full wave function in algorithm II. Instead, we sample the wave function given by
\begin{equation}
   \ket{\psi_0} = \hat{\mathcal{J}}\ket{\phi_0}.
\end{equation}
Since this sampling wave function only contains the reference configuration, its overlap ratios are available in algorithm II. We estimate the energy of the full wave function by taking the ratio of the following quantities obtained by CTMC sampling \(\ket{\psi_0}\):
\begin{equation}
	\begin{split}
   \frac{\expecth{\psi}{H}{\psi}}{\inner{\psi_0}{\psi_0}} &= \sum_n \frac{|\inner{n}{\psi_0}|^2}{\inner{\psi_0}{\psi_0}}\frac{\inner{\psi}{n}}{\inner{\psi_0}{n}}\frac{\expecth{n}{H}{\psi}}{\inner{n}{\psi_0}},\\
	 \frac{\inner{\psi}{\psi}}{\inner{\psi_0}{\psi_0}} &= \sum_n \frac{|\inner{n}{\psi_0}|^2}{\inner{\psi_0}{\psi_0}}\frac{|\inner{\psi}{n}|^2}{|\inner{\psi_0}{n}|^2}.
 \end{split}
\end{equation}
This introduces a bias in the estimate of the energy, which could be severe if \(\ket{\psi_0}\) has vanishing contributions from parts of the Hilbert space where \(\ket{\psi}\) has significant amplitudes.\cite{shi2016infinite} We guard against instabilities due to rare events when the walker overlap with the sampling wave function is small by throwing away outlier samples using a threshold for the magnitude of \(\frac{\inner{\psi}{n}}{\inner{\psi_0}{n}}\). In our experiments, this has not lead to significant issues because \(\ket{\psi_0}\) is a often a good approximation to \(\ket{\psi}\). We note that the occurrences of such rare events can be reduced by including more configurations in the sampling wave function, but we have not done so in this study. This will also improve the sampling efficiency at a slightly enhanced cost introduced by the direct calculation of overlap ratios. We are currently investigating possible sampling issues and will address them more thoroughly in a future article. The gradient can also be sampled similarly.

\section{Results}
In this section, we apply the above formalism to the truncated trans-polyacetylene chain \ce{C28H30}. A model geometry, with uniform bond lengths given by \textit{l}(C=C) = 1.34 \AA, \textit{l}(C-C) = 1.45 \AA, and \textit{l}(C-H) = 1.08 \AA\ and 120 degrees bond angles, was used. We used PySCF\cite{sun2018pyscf} to generate molecular integrals for the 6-31g basis set, and the heat bath CI (HCI)\cite{Holmes2016,sharma2017semistochastic} program Dice to perform an HCISCF\cite{smith2017cheap} calculation with the \(\pi\) active space consisting of 28 electrons in 28 orbitals. Intrinsic bond orbitals\cite{knizia2013intrinsic} obtained by localizing the \(\pi\) orbitals were used in the Jastrow factor. They roughly resemble the \(p_z\) atomic orbitals on each carbon atom.

\begin{figure}[htp]
	\centering
  \includegraphics[width=0.48\textwidth]{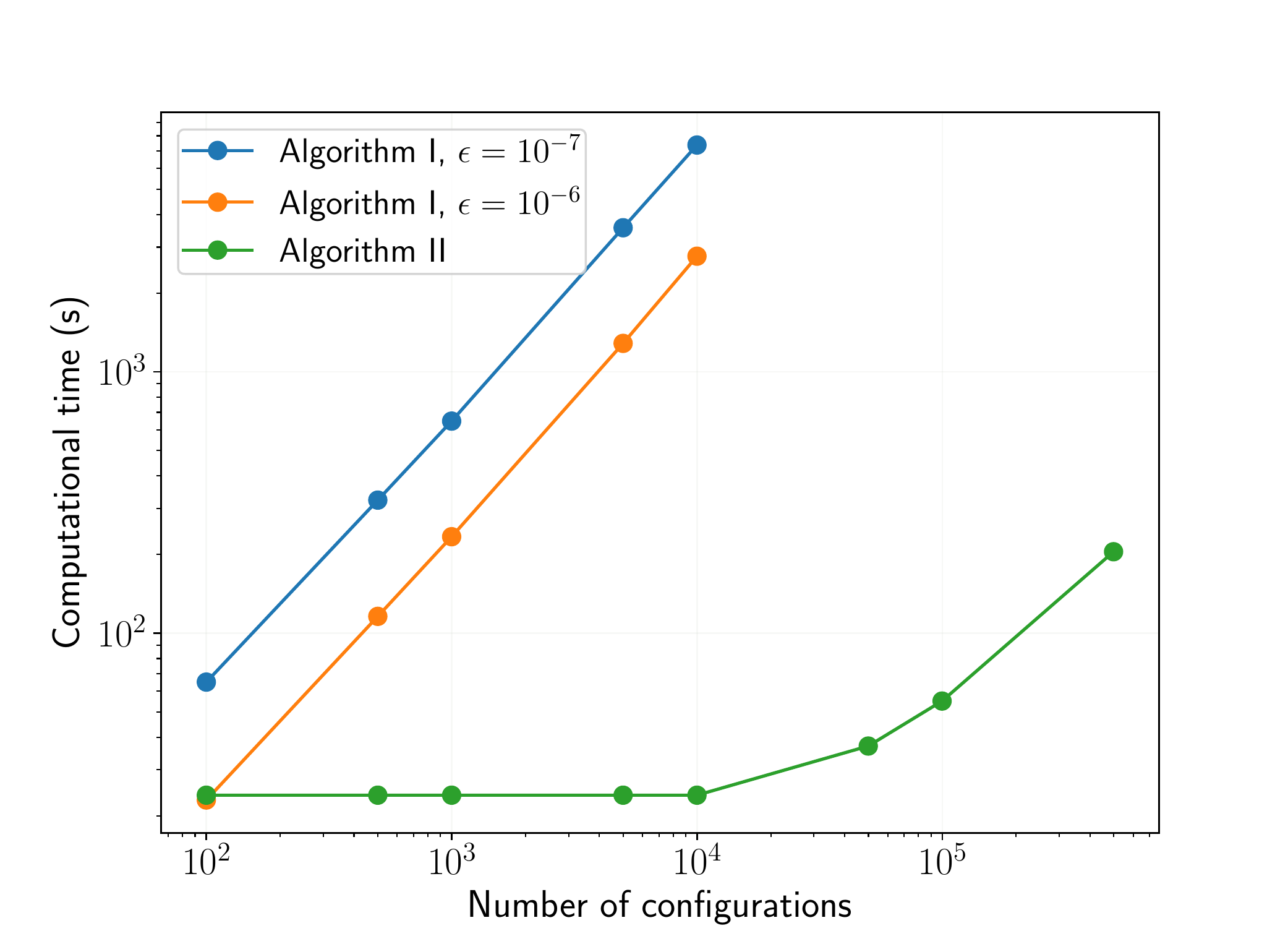}
  \caption{Computational time for calculating 100 local energy samples of the Jastrow multi-Slater wave function plotted against against the number of configurations in the wave function. \(\epsilon\) refers to the Hamiltonian screening threshold.\label{fig:scaling}}
\end{figure}

We first look at the computational performance of the two algorithms. These calculations were performed on an Intel Xeon Gold 6150 2.7 GHz CPU. Figure \ref{fig:scaling} shows the computational cost scaling of the two algorithms as a function of the number of configurations in the wave function, showing the cost of calculating 100 local energy samples during a continuous time Monte Carlo run. The configurations were generated using an SHCI calculation with \(\epsilon_1 = 5\times 10^{-5}\), and only those up to quadruply excited from the reference were retained. These configurations constitute the majority of all leading configurations in the SHCI expansion: of the \(5\times 10^{5}\) leading configurations only 954 were higher than quadruply excited. Algorithm I shows a linear scaling with the number of configurations as expected. We store the Hamiltonian in a heat bath format, which allows efficient screening of the two-electron integrals, where integrals below a screening threshold \(\epsilon\) are ignored. In the figure, we show the cost scaling of algorithm I for two \(\epsilon\) values to demonstrate the effect of screening. Because there is essentially only one orbital on each carbon, screening is very efficient in this case and leads to a significant reduction in the cost of local energy evaluation. But even with aggressive screening, the calculations become very expensive as the number of configurations is increased due to the \(O(n^2n_ck^3)\) scaling.

On the other hand, algorithm II has a much less severe cost scaling. Up to \(1\times 10^4\) configurations the cost is almost independent of the number of configurations. It is dominated by the \(O(n^5)\) scaling calculation of the intermediate in equation \ref{eq:d3}. Because we have implemented this as a dense tensor contraction, screening does not affect the cost. Beyond \(1\times 10^4\) configurations the linear scaling \(O(n_ck^6)\) starts to become dominant. As more quadruply excited configurations are added to the expansion the \(k^6\) factor leads to a change in the slope of the scaling curve. Despite this, the largest calculation with about \(4.9\times10^6\) configurations required 205 seconds for calculating 100 samples with algorithm II, whereas a linear extrapolation of the algorithm I times suggests that it would take more than 38.5 hours for the same calculation even with aggressive screening. This analysis demonstrates the favorable scaling of algorithm II, making local energy calculations with long HCI expansions feasible.

\begin{figure}[htp]
	\centering
  \includegraphics[width=0.48\textwidth]{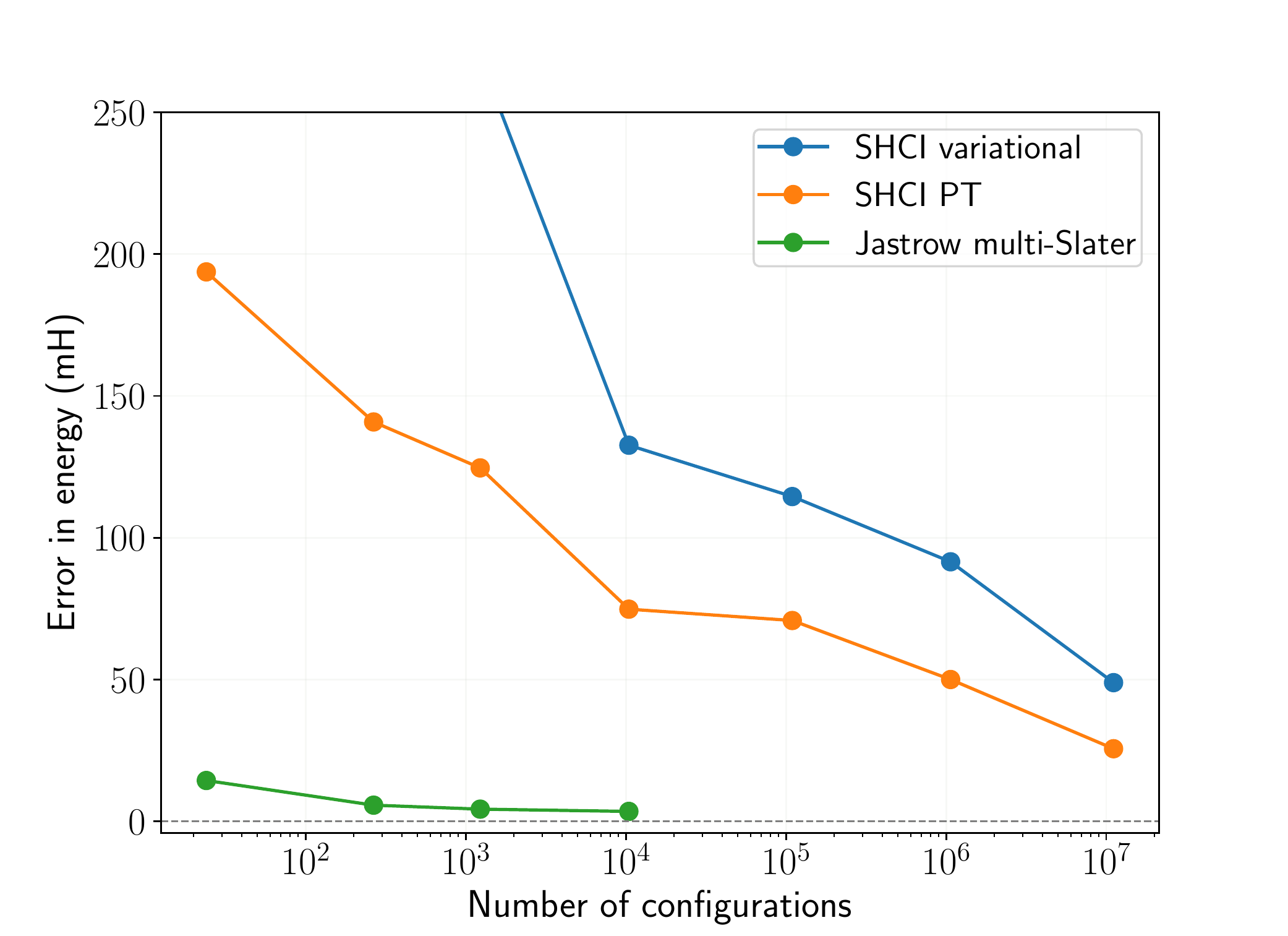}
  \caption{Convergence of ground state energy with the number of configurations for SHCI variational and perturbation theory energies, as well as for the Jastrow multi-Slater wave function. Statistical errors are smaller than the symbol size.\label{fig:ene}}
\end{figure}

In figure \ref{fig:ene}, we show the effect of the Jastrow factor on the convergence of multi-Slater ground state energy. The reference energy was obtained by performing a density matrix renormalization group calculation, which is very accurate for linear systems. SHCI calculations were performed with progressively smaller \(\epsilon_1\) values leading to progressively longer expansions. Even after including the perturbation theory correction, SHCI has significant difficulty in converging to the exact energy because of the large number of strongly correlated degrees of freedom in this active space. The ground state is dense in the canonical representation, and has significant contributions from highly excited configurations, leading to slow convergence of SHCI.

For the Jastrow multi-Slater wave function, we used leading configurations from an SHCI calculation performed with \(\epsilon_1 = 5\times 10^{-5}\). Note that these configurations are different from those included in the SHCI wave functions used to examine the convergence of SHCI energies.
We made this particular choice to make sure that wave functions with more configurations are variationally superior than those with fewer. This choice also allowed us to use the parameters optimized for the shorter expansions to be used as initial guesses for longer ones, which was crucial for the optimization. We used stochastic gradient descent with momentum\cite{goh2017momentum} to optimize the wave function energies. Jastrow parameters and CI coefficients were optimized together. We used 84 processes with 100 CTMC samples and 10 burn-in samples each during the optimization and did a final energy calculation with 3000 samples per process to get the statistical error below 0.5 mH. The Jastrow multi-Slater energies can be seen to converge much more rapidly than bare SHCI. The energy error for the wave function consisting of about \(1.0\times 10^4\) configurations is 3.5(4) mH, whereas the error in the HCI variational energy with \(1.1\times10^7\) configurations is 48.8 mH. Similar observations have been made for the Hubbard model using transcorrelated Hamiltonians.\cite{tsuneyuki2008transcorrelated,dobrautz2019compact}

\section{Conclusions}
We have presented an algorithm for calculating local energies of multi-Slater wave functions in orbital space Monte Carlo. We used a Jastrow multi-Slater wave function in VMC to demonstrate its efficiency. The algorithm itself is general and should allow the use of longer selected CI expansions as trial wave functions in other orbitals space QMC methods like AFQMC and Green's function Monte Carlo\cite{van1994fixed,Haaf1995} as well. It also presents a different formulation of the real space algorithm proposed by Filippi \textit{et al.} for local energies. We also showed the efficacy of Jastrow factors in vastly improving the convergence of the energy of CI expansions in polyacetylene.

We plan to explore various possibilities for improving the algorithm and using it in other Monte Carlo methods. As mentioned before, methods like density fitting and tensor hypercontraction can be used to reduce the cost of building the intermediates, which will likely dominate the total cost for large systems. We would also like to explore other ways of exploiting the structure in the Hamiltonian. Our pilot implementation uses dense tensor contractions even for the sparse Hamiltonian, which can be improved considerably. As for the Jastrow multi-Slater wave functions, we would like to study the criteria for choosing configurations in the presence of a Jastrow, as well as the possibility of using symmetry projection to improve the convergence of energy with the number of configurations.\cite{tahara2008variational,scuseria2011projected} This will be crucial for feasibly optimizing these wave functions because stochastic optimization with a large number of parameters is a difficult task. Our numerical experiments with polyacetylene suggest that an optimization strategy where CI coefficients are progressively optimized in blocks may be more effective than optimizing all of them together. Based on experiences in the community with other nonlinearly parameterized wave functions, it may be worth exploring such optimization techniques.

\begin{acknowledgments}
The funding for this project was provided by the national science foundation through the grant CHE-1800584. SS was also partly supported through the Sloan research fellowship.
\end{acknowledgments}

\section*{Data availability statement}
The data that support the findings of this study and the code used to generate them are available from the authors upon reasonable request.


%


\providecommand{\latin}[1]{#1}
\makeatletter
\providecommand{\doi}
  {\begingroup\let\do\@makeother\dospecials
  \catcode`\{=1 \catcode`\}=2 \doi@aux}
\providecommand{\doi@aux}[1]{\endgroup\texttt{#1}}
\makeatother
\providecommand*\mcitethebibliography{\thebibliography}
\csname @ifundefined\endcsname{endmcitethebibliography}
  {\let\endmcitethebibliography\endthebibliography}{}

\end{document}